\RequirePackage{ifpdf}
\documentclass{JHEP3}
\pdfoutput=1

\usepackage{rotating}
\usepackage{graphicx}
\usepackage{subfigure}
\usepackage{amssymb}
\usepackage{amsmath}
\usepackage{amsfonts}
\usepackage{latexsym}
\usepackage{url}
\usepackage{enumerate}
\usepackage{cite}
\usepackage{float}

\newcommand{\IP}{\textrm{IP}}
\newcommand{\IPS}{\textrm{IPS}}
\newcommand{\trk}{\textrm{trk}}

\title{Data Driven Search in the Displaced $b\bar{b}$ Pair Channel for a Higgs Boson Decaying to Long-Lived Neutral Particles}
\author{Valerie Halyo,$^a$ Hou Keong Lou,$^a$ Paul Lujan$^a$, Wenhan Zhu$^a$\\
$^a$ Department of Physics, Princeton University, Princeton, NJ 08544, USA\\
\email{valerieh@princeton.edu}\\
\email{hlou@princeton.edu}\\
\email{plujan@princeton.edu}\\
\email{wenhanz@princeton.edu}
}
\abstract{
  This article presents a proposal for a new search channel for the
  Higgs boson decaying to two long-lived neutral particles, each of
  which decays to $b\bar b$ at a displaced vertex. The decay length
  considered is such that the decay takes place within the LHC
  beampipe. We present a new data-driven analysis using jet substructure
  and properties of the tracks from the highly-displaced vertices. 
  We consider a model with a 125 GeV Higgs boson with a significant branching fraction to decay via this mode,
  with the long-lived neutral particle having a mass in the range of 15--40 GeV and a decay length
  commensurate with the beam pipe radius.
  Such a signal can be readily observed  with an integrated
  luminosity of 19.5 fb$^{-1}$ at 8~TeV at the LHC.
}
\preprint{MAD}

\begin{document}

\section{Introduction}
\label{sec:intro}

Until recently, the Higgs sector was one of the great unknowns in our current understanding of particle
physics, and the primary target of the current Tevatron and Large Hadron Collider (LHC) programs. The new
scalar boson recently discovered at the LHC by the ATLAS and CMS
experiments~\cite{Aad:2012tfa,Chatrchyan:2012ufa} has, so far, been measured to be consistent with a Standard
Model (SM) Higgs boson~\cite{ATLAS:2013sla,CMS:yva}. However, the systematic uncertainties in the measurements
still allow for the possibility that this new particle could be responsible for electroweak symmetry breaking
and mass generation but not be the SM Higgs boson.  In particular, it could have non-SM properties such as
mixing with a singlet, non-standard couplings to the fermions, or other exotic decays. 
If one assumes a Higgs with SM couplings except additional decay channel to new particles, a branching ratio as large as 20\% can be accommodated given the 2012 LHC data \cite{Belanger:2013kya,Ellis:2013lra}.

In this article we address an exotic Higgs decay mode that would have escaped existing search strategies.  We
consider the possibility \cite{Strassler:2006ri} (see also \cite{Chang:2005ht,Carpenter:2007zz} for closely
related work) that the Higgs boson $h$ decays to two spin-zero neutral particles $X$, and the $X$ decays in
turn to $b\bar b$ with a displaced vertex.  More specifically, we will consider the case where the lifetime
$\tau_X$ of the $X$ puts its decay at a distance from the collision point of order millimeters to a few
centimeters, so that the decay vertex remains within the LHC beampipe. Searches for related signatures have
also been made in D0 and ATLAS. In D0, the typical decay to two $b\bar{b}$ pairs in the several to 20
centimeter range has been studied and weakly constrained~\cite{Abazov:2009ik}. In ATLAS, strong limits on
final states with a muon and multiple displaced jets have been obtained~\cite{Aad:2012zx}; however, as the
model considered involves an $R$-parity-violating neutralino decay into a muon and hadrons, the transverse
momentum of the muon was required to be higher than 50 GeV/$c$, which is unlikely to result from the
semileptonic decay of the bottom quark used in our model.

It is sometimes argued that searches of this type are not so well-motivated, because the chance of the $X$
having a lifetime that allows for decays inside the detector is low.  However, there are both theoretical and
experimental considerations in favor. First, long-lived particles are less rare in models
~\cite{Barbier:2004ez,Dermisek:2005ar,Chang:2005ht,Strassler:2006ri,Carpenter:2007zz,AristizabalSierra:2008ye,Kang:2008ea,Cheung:2008ke,Bellazzini:2009xt,Juknevich:2009ji,Falkowski:2010cm,Englert:2011us,Graham:2012th}
than is commonly assumed.  In hidden valley models (\cite{SZhv}, for instance), there may be not one but many
new particle states with a wide variety of lifetimes, similar to the case of QCD, and this plenitude makes it
more likely that one of these particles will have a detectable displaced decay. Second, decays of such
particles have such limited SM background that in principle only a few such events might suffice for a
discovery, so even a small branching fraction for such particles may lead to a discovery opportunity.  That
said, detector backgrounds can be a serious issue, and event triggering and reconstruction may be an even
larger one if the lifetimes are long enough.  Each search strategy has its own features, and some are easier
than others.

The Tevatron and LHC detectors were generally not optimized for finding long-lived particles, with the exception of $B$ hadrons, 
and searches for such particles face numerous challenges.  In this paper we consider the case that, relatively speaking, is the 
easiest: a search for a new particle that mainly decays before that particle reaches the beampipe.  Such decays face little or no 
background from secondary interactions of hadrons with detector material, and the dominant background is a physics background from 
real $B$ hadron decays.   However, to the extent the $X$ lives longer than the $B$ hadron and is considerably heavier, distinguishing 
it from SM heavy-flavor backgrounds should be easier.  For the specific case of $h\to XX$, the situation is better still, since there
 are two $X$ decays per event, and also a mass resonance that may be reconstructable.

The main purpose of our paper is to suggest a search strategy for $h\to XX$, with $X$ decaying to $b\bar{b}$
before passing through the wall of the beampipe.  These events are selected online with a trigger requiring a
single muon and two $b$-jets tagged using an algorithm measuring the secondary vertex displacement. When the
mass of the $X$ is heavy, the resulting $b$-jets typically have low $p_{T}$ and cannot be triggered
efficiently. Hence, we focus on the region where $X$ is light and so the jets from the two $b$-quarks merge
into a single reconstructed jet. Also, as we will describe in this paper, by merging two $b$-quarks into one
jet, the QCD background can be estimated using data-driven methods. Data-driven techqniues are crucial for low
mass signals, where systematic uncertainty can be very large. To extract the signal, we then select $X$ boson
candidates by looking for jets meeting a combination of two requirements: first, using the displacement of the
individual tracks to identify long-lived particles, and second, using the internal substructure of the jet to
further distinguish exotic displaced jets from displaced $b$-jets. By combining jet substructure with
displaced tracks and vertices, it is possible to devise a new exotic jet tagger and propose a data-driven
method for estimating the QCD background.  Using our technique, it is demonstrated that new long-lived neutral
particles originating from a 125 GeV/$c^2$ Higgs boson may be discovered using 19.5 fb$^{-1}$ of LHC data
recorded at $\sqrt{s} = 8$ TeV.

For our studies, we focus on a model with a non-SM Higgs with a mass of 125 GeV/$c^2$, which then decays into a
pair of long-lived neutral bosons $X$ whose mass ranges from 15 to 40 GeV/$c^2$. For the boson $X$, we
primarily consider the case where the $X$ subsequently decays into $b\bar{b}$. Because of the relatively
low mass of the $X$ bosons considered, the $b\bar{b}$ pair is generally reconstructed as a single jet in
the detector. We use the anti-$k_T$~\cite{Cacciari:2008gp} algorithm with $\Delta R=1.0$ in this
analysis to capture the hadrons from both $b$ quarks in a single object, which we refer to as a ``fat
jet''.  The final topology consists of two fat jets producing a resonance at the expected Higgs mass,
with a distinctive two-prong jet substructure in each jet. The predominant background is the QCD
production of $b\bar{b}$ pairs.  However, since these pairs are being produced from a single quark, they
will tend to have fewer displaced tracks, and will tend to contain only one central hard prong. These
properties allow us to differentiate the signal from the much larger background.

\begin{figure}[ht]
\begin{center}
  \subfigure[]{\includegraphics[width=0.45\textwidth]{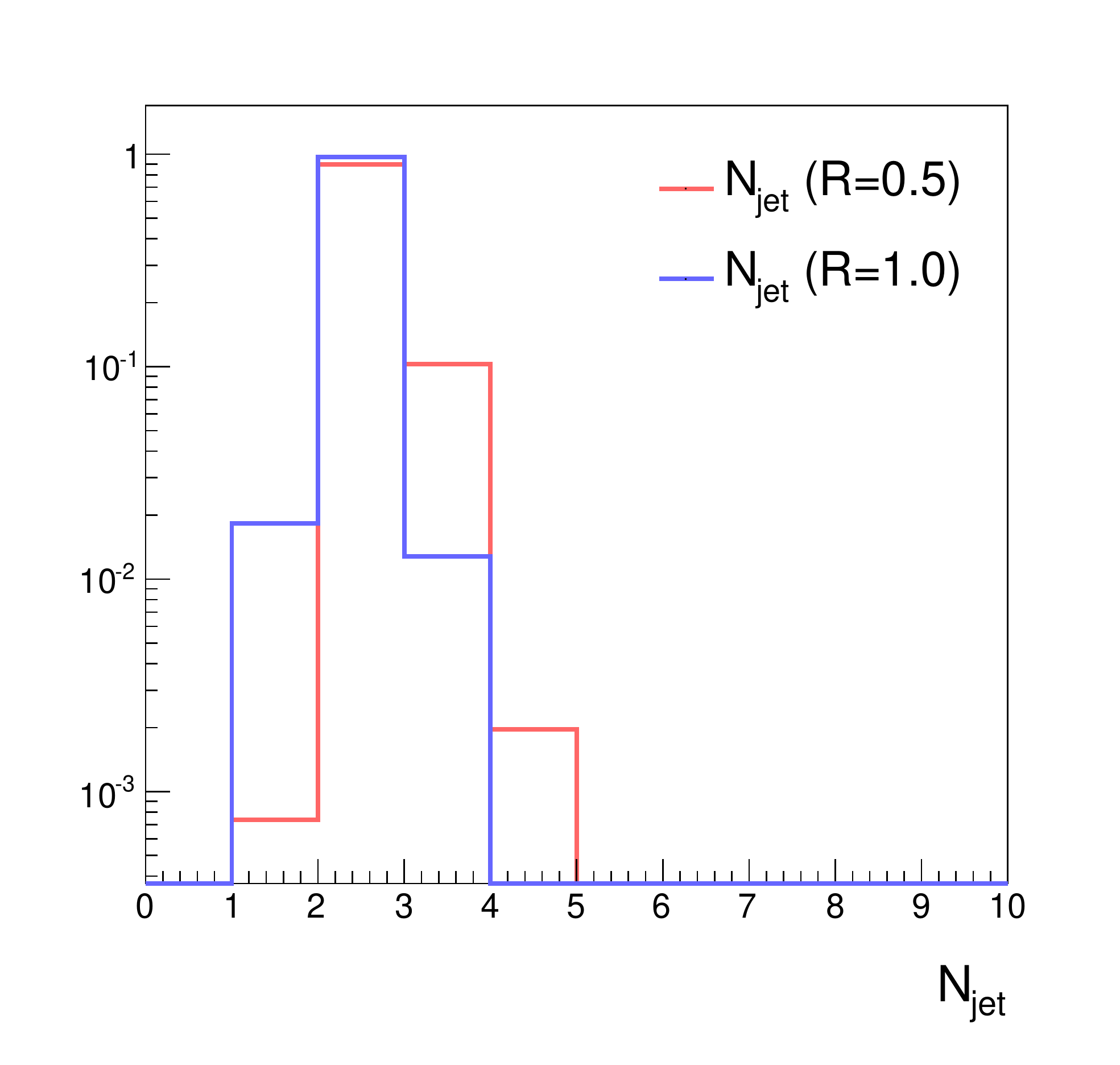} \label{fig:signal_njet}}
  \subfigure[]{\includegraphics[width=0.45\textwidth]{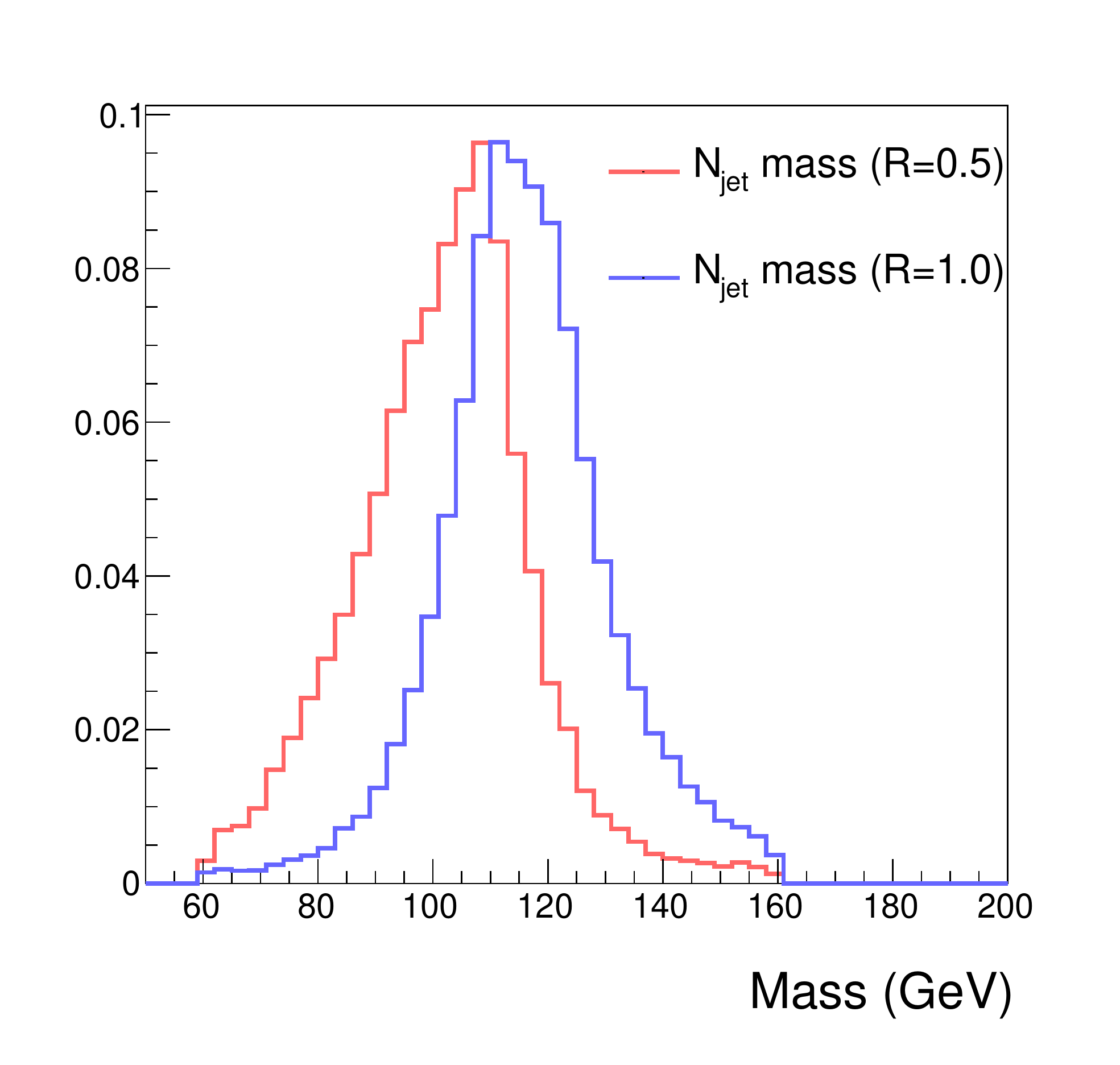} \label{fig:signal_mass}}
  \caption{Results of jet clustering in simulated signal samples. Left: number of reconstructed jets using the
    standard cone radius of $\Delta R = 0.5$ (red) and our cone radius of $\Delta R = 1.0$ (blue). Right:
    Reconstructed two-jet mass for these two different cone radii. The signal model shown is with $m_H=125$
    GeV/$c^2$, $m_X=20$ GeV/$c^2$, and $c\tau=2$ mm.}
  \label{fig:signal_jets}
\end{center}
\end{figure}

The need for using large-radius jets instead of the standard cone size of $\Delta R = 0.5$ is illustrated in
Figure~\ref{fig:signal_jets}. This figure shows the results of jet reconstruction in a simulated signal sample
with $m_H=125$ GeV/$c^2$, $m_X=20$ GeV/$c^2$, and $c\tau=2$ mm for two different cone sizes: the standard cone
size of $\Delta R = 0.5$, and our enlarged cone size of $\Delta R = 1.0$. In Figure~\ref{fig:signal_njet}, we
see that even with the standard cone size, in the vast majority of events the $X \rightarrow b\bar{b}$ decay
is reconstructed as a single merged jet, rather than two separate jets. However, Figure~\ref{fig:signal_mass}
shows that the standard cone size is too small to capture all of the radiation from this merged jet, resulting
in a significant underestimation of the reconstructed mass. Using a larger cone radius thus offers two
advantages: first, the event is nearly always reconstructed with exactly two jets, allowing for more
predictable reconstruction; and second, the cone size is large enough to capture all of the merged jet,
allowing for more accurate mass reconstruction. We can then use subjet techniques on the merged jets to
identify the two constituents.

In this article a variety of proper lifetimes $c\tau$ for the long-lived particle ranging from 1~mm to
10~mm was considered. At very low lifetimes, the displacements of the resulting tracks and vertices become
too small to consistently separate the signal events from background, while at high lifetimes the track
and muon reconstructions suffer from inefficiencies in the tracking and trigger algorithms, which are not
generally designed for highly-displaced particles. However, our simple detector simulation will be unable
to take these effects into account and hence we refrain from extending our results beyond 30~mm.

Although our search strategy uses $h\to XX$ as a benchmark for
optimization, it is not strongly dependent on the specific initial or
final state.  Consequently it should be somewhat model-independent,
and would be sensitive to a variety of models with two long-lived
particles in the events. For example, certain gauge-mediated
supersymmetric models with a neutralino~\cite{MatchevThomas} decaying
in flight to a $Z$ or $h$ might be picked up by our search.  One point
of model-dependence worth keeping in mind is that the heavy-flavor
content of the $X$ decay is important for our strategy, as we will
base our study on a $b$-tagger-like trigger.

\section{Event Generation}
\label{sec:generation}
At hadron colliders the dominant Higgs production mechanism is via gluon-gluon fusion. In this note we study
the process $gg\rightarrow h\rightarrow(X\rightarrow b\bar{b})(X\rightarrow b\bar{b})$, where the Higgs boson
is produced by gluon fusion and then decays into a pair of long lived (pseudo-)scalars which then each decay
to a pair of bottom quarks. We consider this in the context of $pp$ collisions at a center-of-mass energy of
$8$~TeV.

We generate the signal samples for Higgs mass $m_h=$ 125 GeV/$c^2$ and the \linebreak (pseudo-)scalar mass
$m_X$ between 15 and 40 GeV/$c^2$ with 5 GeV/$c^2$ steps. Samples were generated for the $c\tau_X$ of the
scalar varying within a wide range between 0.1 mm to 30 mm.  The signal sample is generated using Pythia
6.4.27~\cite{Sjostrand:2006za}. For the production cross section, we use the NLO cross section for SM $gg
\rightarrow h$ production, which is 19.3~pb at 8 TeV~\cite{Heinemeyer:2013tqa}.

The dominant background for this process comes from QCD heavy quark production, particularly events with one
or more $b\bar{b}$ pairs, which represents the most difficult background to remove. Using \textsc{MadGraph 5}
v1.5.7~\cite{Alwall:2011uj}, we generated a sample of 500 million $b\bar{b}$ events matched up to four jets
(including $b\bar{b}b\bar{b}$ and $b\bar{b}c\bar{c}$), and showered them through
\textsc{Pythia}~\cite{Sjostrand:2006za}.  Matching is done using the MLM prescription \cite{Mangano:2006rw}.
In order to account for effects that may not be fully modeled in the simulation, a K-factor of 1.6 is obtained by
generating another QCD $b\bar{b}$ sample at 7 TeV and reproducing the CMS analysis published in
\cite{Chatrchyan:2013qga} (see Appendix \ref{sec:validation} for details and discussion on effects from fake
$b$-tags). We used the CTEQ6L1 PDF for both the signal and the background \cite{Pumplin:2002vw}.

%
%

To simulate particle flow jets at CMS, the stable particles (except neutrinos) are clustered into large anti-$k_{T}$~\cite{Cacciari:2008gp} jets
with a cone size of $\Delta R=1.0$ using \texttt{FastJet} 3.0.2~\cite{Cacciari:2011ma}. Because our jets use an anti-$k_{T}$ algorithm with a large cone size to capture as much of the $b\bar{b}$ decay as possible, they are more susceptible to underlying event and pileup
effects. In order to overcome these issues, we use jet trimming with $R_{\textrm{trim}}=0.3$ and
$f_{\textrm{cut}}=0.05$ \cite{Krohn:2009th}. The resulting jets are then smeared
using the momentum resolution given in~\cite{CMS:2009nxa}~\footnote{The values for jets with a cone size
$\Delta R=0.5$ are used; however, as the measured momentum is relatively unimportant to our analysis, this
difference should not have a significant effect.}. To simulate the detector response in the tracker, we associate a track to each charged final state particle with $p_{T} > 1$ GeV/$c$. The production vertex of the track is then smeared by an uncertainty $\sigma_{\trk}$,
extracted from \cite{CMS:2010wta}:
\begin{align}
\label{eqn:sigma}
\sigma_{\trk}&=a+\frac{b}{p_{T}}+\left(c+\frac{d}{p_{T}}\right)\eta^2 \\
a=20.4 \quad b=&56.4 \quad c=-0.11 \quad d=18.2, \notag
\end{align}
where $\sigma_{\trk}$ is in units of $\mu$m and $p_{T}$ in units of GeV/$c$. To avoid complications in finding
the vertex location of our event along the beamline, we do not use any tracking information in the longitudinal
direction. The QCD background is validated against published CMS
results. For details, see Appendix \ref{sec:validation}. 

\section{Displaced Jet Variables}
\label{sec:djet}
In the following, we discuss the key variables used for identifying long-lived decays. We postpone the discussion on event
selection until Section \ref{sec:selection}, where all the analysis strategies and cuts are listed in detail.

\begin{figure}[ht]
\centering
\includegraphics[width=0.45\textwidth]{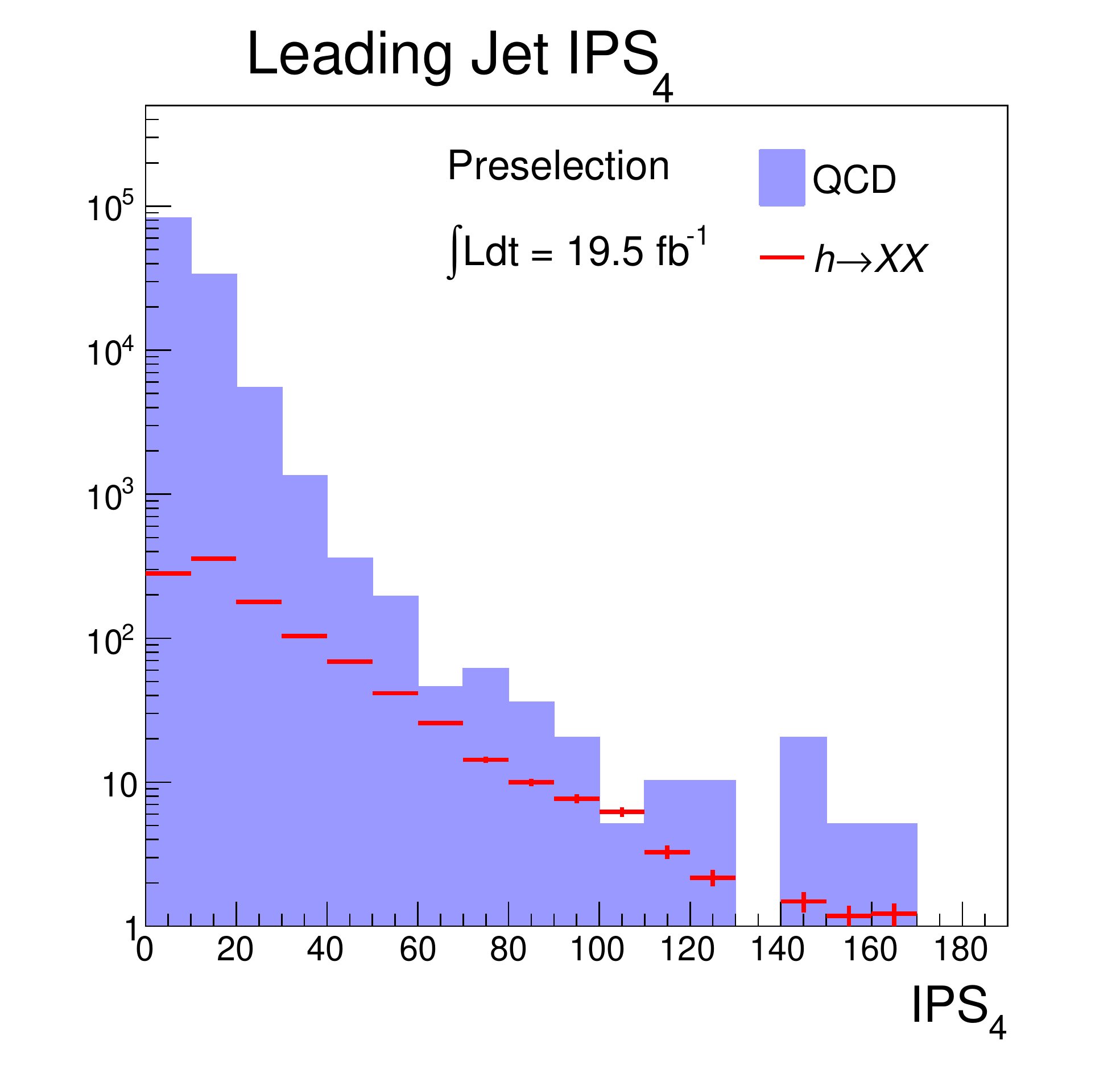}
\includegraphics[width=0.45\textwidth]{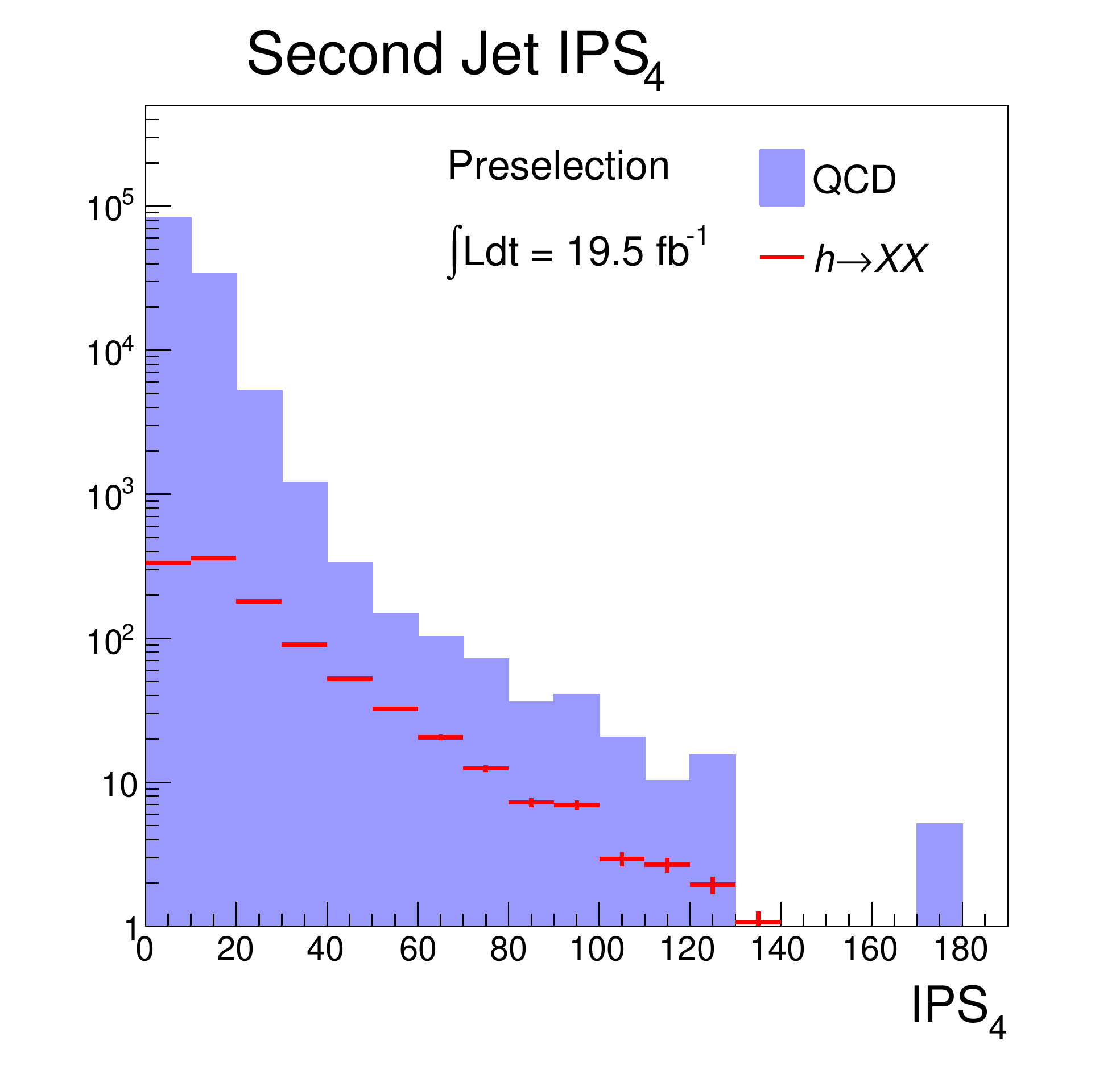}
\caption{Distributions of $\IPS_{4}$ after preselection. The QCD background is given by the matched $b\bar{b}$
sample, and the signal model shown is with $m_H=125$ GeV/$c^2$, $m_X=20$ GeV/$c^2$, and $c\tau=2$ mm. The signal is
assumed to have SM gluon fusion production with 100\% branching ratio to $XX$. Left: for the leading jet. Right: for
the second-leading jet.}
\label{fig:IPS4}
\end{figure}

The primary tool we use to measure the displacement of a jet is to examine the displacement of the individual
tracks in the jet. For each track, we compute the transverse Impact Parameter (IP) as follows:
\begin{align}
\IP=\frac{|v_x\cdot p_y - v_y\cdot p_x|}{p_{T}}
\end{align}
The computed $\IP$ has an associated error $\sigma_{\IP}^2=\sigma^2_{\trk}+ \sigma^2_{\textrm{PV}}$, where
$\sigma_{\trk}$ is given by equation~\ref{eqn:sigma}, and $\sigma_{\textrm{PV}}=0.025\,\mu\textrm{m}$ is the uncertainty associated
with the transverse coordinate of the Primary Vertex (PV), as determined in \cite{CMS:2010wta}.  In an actual
detector, the track IP errors are often as large as the measured track IP itself, and it is advantageous to
consider the impact parameter significance\footnote{Our $\IPS$ distribution shows good agreement when compared
to the CMS results shown in Figure 3 of \cite{Rizzi:2006ms}. For details see Appendix \ref{sec:validation}.} (IPS) \cite{Chatrchyan:2012jua}

\begin{align}
\IPS=\frac{\IP}{\sigma_{\IP}}
\end{align}
For prompt tracks,  the $\IPS$ distribution tends to have a strong peak around zero and the
spread of the distribution depends on the mismeasurement of the $\IPS$ or misalignment. 
For genuine displaced tracks, the $\IPS$ distribution will tend to have a
significant tail. Validation of the $\IPS$ variable against published data is presented in Appendix \ref{sec:validation}.

For each jet, we order the associated tracks in decreasing $\IPS$. We then consider the
fourth-highest $\IPS$ value, denoted by $\IPS_{4}$. A typical $b$-jet will tend to only have two displaced
tracks, while a displaced $b\bar{b}$ pair will have four, so this variable is expected to have significant
discriminating power. Figure~\ref{fig:IPS4} shows the $\IPS_{4}$ distributions for signal vs. QCD.

Other variables, such the significance of the decay length of the jet vertex and the fraction of the jet energy
carried by prompt tracks, were considered as discriminants for identifying displaced particle decays.
However, they did not result in a significant increase in discriminating power, so due to the lack of
validation of these other variables and for the sake of simplicity,  we do not use these in our analysis.

\section{Jet Substructure}
\label{sec:substructure}

\begin{figure}[ht]
\centering
\includegraphics[width=0.45\textwidth]{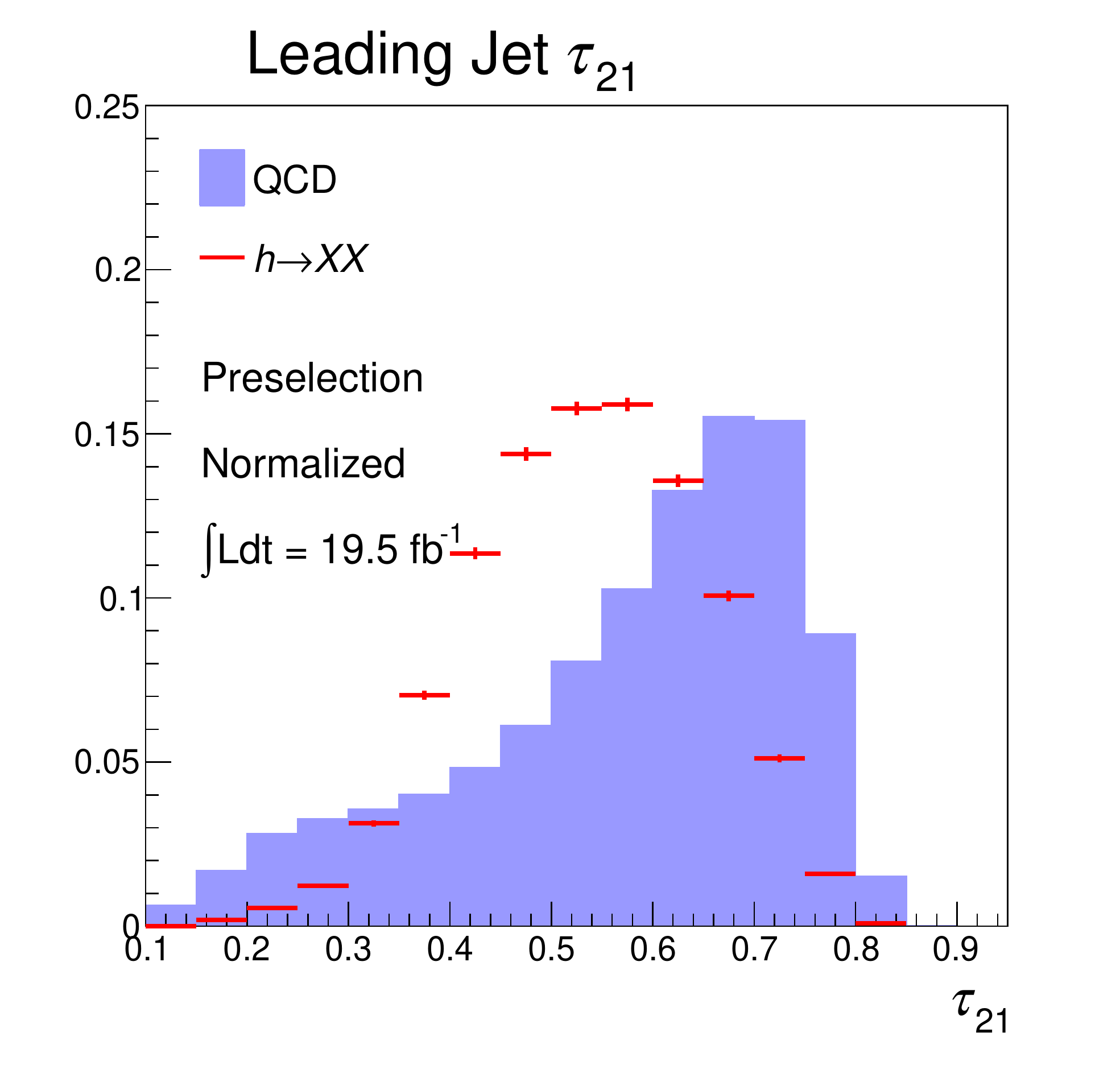}
\includegraphics[width=0.45\textwidth]{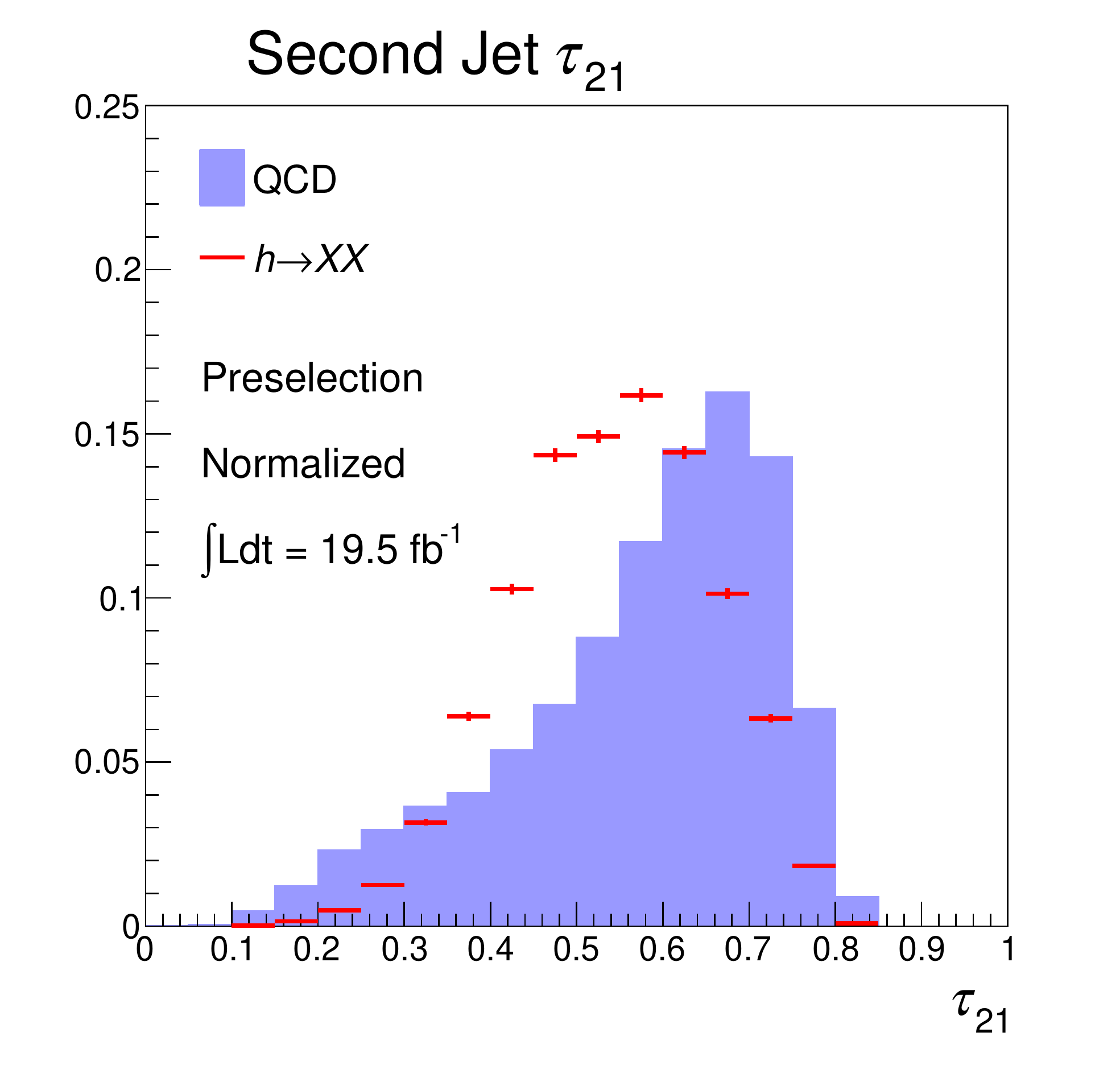}
\caption{Distribution of $\tau_{21}$ for the leading jet (left) and the second-leading jet (right) for
  simulated signal events and $b\bar{b}$ background, for a signal model with $m_H = 125$ GeV/$c^2$ and $m_X =
  20$ GeV/$c^2$.}
\label{fig:tau21}
\end{figure}

In this analysis we look for ``fat jets'' that originate from the decay of the long-lived particle into a
$b\bar{b}$ pair, and we expect the ``fat jet'' to contain a different substructure than jets originating from
a single $b$ quark. In order to quantify this substructure, we use the ``$N$-subjettiness'' variables defined
in~\cite{Thaler:2010tr,Thaler:2011gf}. Briefly, one first defines $\tau_N$ by fitting $N$ axes to a jet, and
computing
\begin{equation}
\tau_N = \frac{1}{d_0}\sum_{k}p_{T,k} \min\{\Delta R_{1,k}, \ldots \Delta R_{N,k}\},
\end{equation}
where $k$ runs over the constituents in the jet and $d_0$ is an unimportant
overall normalization factor. $\tau_N$ is then minimized over all possible choices of
the $N$ subjet axes. $\tau_N$ thus shows to what degree the jet can be viewed as being composed of $N$ individual
subjets. For distinguishing jets with two subjets from one, we use $\tau_{21} = \tau_2 / \tau_1$. If
$\tau_{21}$ is close to 0, that indicates that the jet is strongly favored to have two subjets, as we would
expect from our signal jets, while a $\tau_{21}$ close to 1 indicates that the jet does not have a two-subjet
structure, as we would expect from QCD. One can see from Figure~\ref{fig:tau21} that the $\tau_{21}$
distributions are indeed different between signal and QCD\footnote{To combat underlying event and pile-up, jet trimming is applied using $R_{\textrm{trim}}=0.3$ and $f_{\textrm{cut}}=0.05$}.

\section{Event Selection}
\label{sec:selection}

Following the definition of variables of interest to this analysis, we describe the selection criteria devised
to select events containing $X$ boson candidates.

\subsection{Trigger}

We simulate one of the High Level Triggers (HLT, purely software-based and with access to the full event
information) used in a CMS Higgs search \cite{Chatrchyan:2013qga}.  As it is difficult to achieve a good
trigger efficiency and purity with a purely hadronic trigger, we instead focus on events where at least
one of the $b$ quarks decays semileptonically to a muon. Thus, we require the events to contain at least one
muon with $p_{T}>$ 12 GeV/$c$ and two $b$-tagged jets with $p_{T} >$ 40 GeV/$c$ and 20 GeV/$c$, respectively. Our
simulation adopts a slightly simpler method for $b$-tagging than that used online in CMS, and uses a
track counting method \cite{Chatrchyan:2012jua} which requires a $b$-tagged jet to have at least two
tracks with $\IPS >$ 3. This custom selection offers a $\sim$50 \% efficiency, comparable to that from CMS. The trigger is fully efficient after preselection.

\subsection{Selection for $X\rightarrow b\bar{b}$}

We expect the signature of the $h \rightarrow XX \rightarrow 4b$ event to be two displaced fat jets, where each
displaced fat jet originates from a $b\bar{b}$ pair, and at least one muon produced 
by the semileptonic decay of a $b$ quark in the event. 
The selection is applied in two stages. First, we apply a preselection with relatively loose requirements
on the displacement of the jets; the primary purpose of the preselection is to eliminate light-flavor
background so that only signal and $b\bar{b}$ background remains. The preselection also reduces the
correlation between the two jets, allowing us to treat them as uncorrelated. After the preselection is
applied, a final selection, using the displacement and the jet substructure, is used to separate the
signal from the $b\bar{b}$ background.
The various selections, applied sequentially, are described below, and the yields for
signal and background are presented in Table~\ref{tab:preselection}. The preselection consists of the following four requirements:

\begin{enumerate}[Cut 1:]
\item The event must pass the simulated trigger, as described above.
\item At least two fat jets, constructed as described previously, satisfying the following quality requirements:
  \begin{itemize}
    \item $|\eta| < 3.0$ and $p_T >$ 30 GeV/$c$
    \item At least 8 associated tracks per jet
  \end{itemize}
\item One of the two chosen jets must match to a muon with $p_T > 12$ GeV/$c$.
\item Both jets must have $\IPS_{4} > 5$.
\end{enumerate}

\begin{figure}[ht]
\centering
\includegraphics[width=0.45\textwidth]{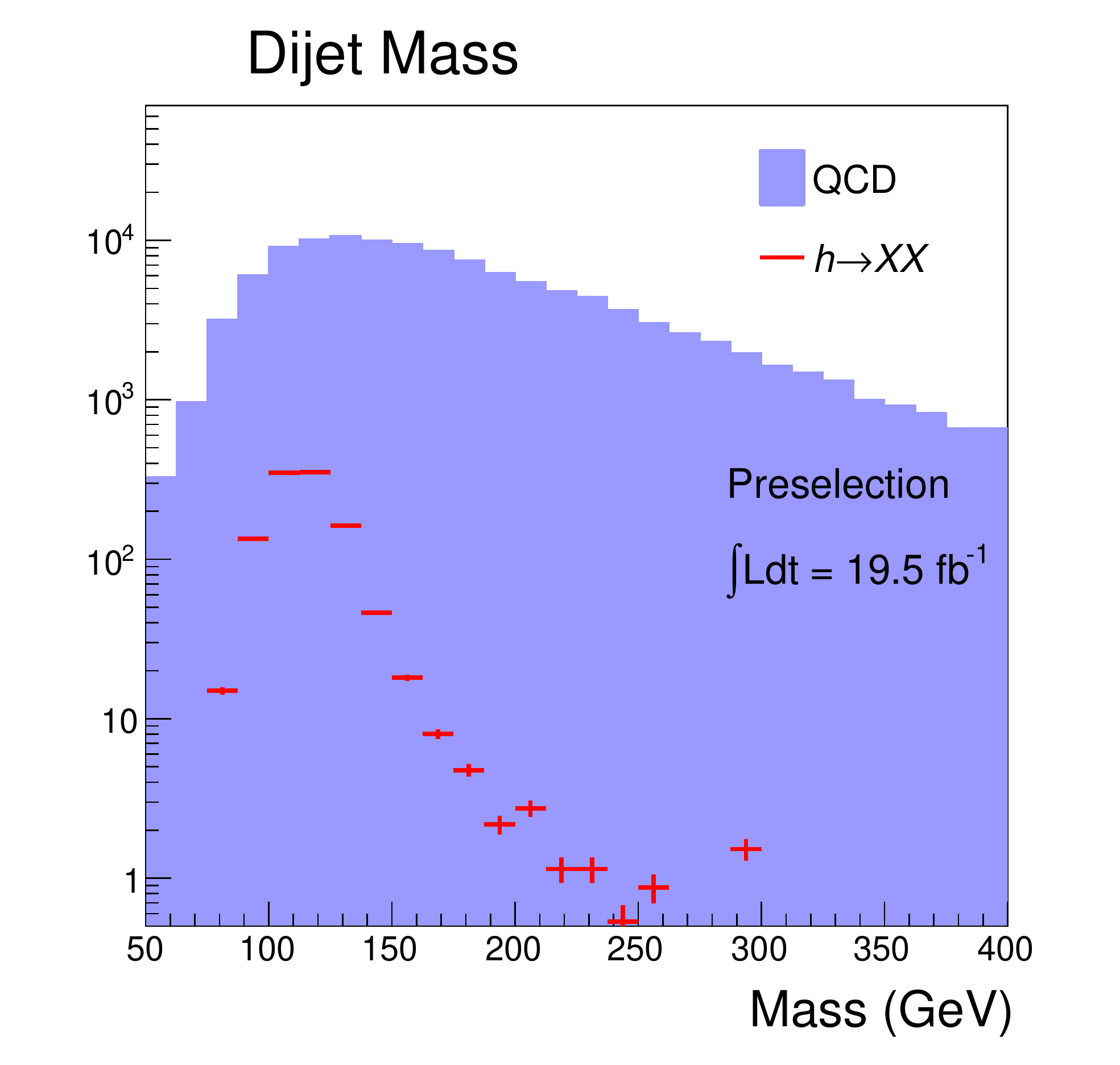}
\includegraphics[width=0.45\textwidth]{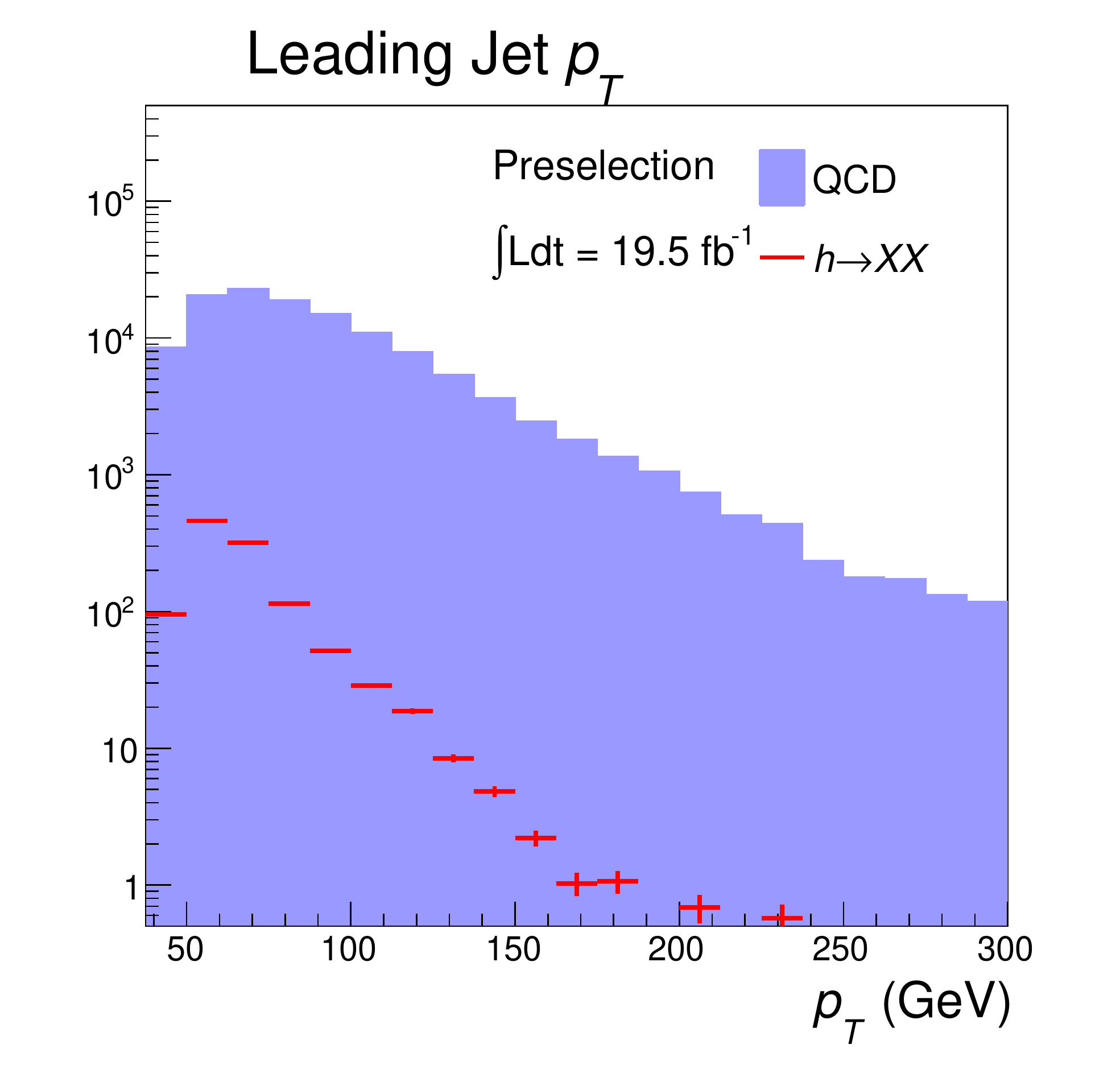}
\caption{Distributions of some kinematic quantities after preselection. Left: $m_{jj}$, the invariant
  mass of the two leading jets in the event. Right: $p_T$ of the leading jet in the event. Shown here are
  simulated signal events and $b\bar{b}$ background, for a signal model with $m_H = 125$ GeV/$c^2$ and $m_X =
  20$ GeV/$c^2$.}
\label{fig:preselection}
\end{figure}

After preselection, the leading two fat jets are essentially determined to be real $b$-jets. This reduces the
correlation between the two leading jets, which is crucial for the data-driven analysis. Figure~\ref{fig:preselection} shows some distributions of the jets after this preselection is
applied. Table~\ref{tab:preselection} shows the expected efficiency of these cuts in the background and
signal simulation.

\begin{table}[h]
\begin{center}
\begin{tabular}{|r|c|c|c|c|}
\hline
 & \multicolumn{2}{c}{Background} & \multicolumn{2}{c|}{Signal} \\
\hline
Cut & Number of events & Efficiency (\%) & Number of events & Efficiency (\%) \\
\hline
Trigger & 7.4 x $10^7$ & --- & 1.6 x $10^4$ & --- \\
Jet quality & 1.2 x $10^7$ & 15.8 & 6.7 x $10^3$ & 42.4 \\
Muon match & 9.1 x $10^6$ & 78.2 & 4.7 x $10^3$ & 70.5 \\
$\IPS_{4}$ & 2.8 x $10^5$ & 3.1 & 1.4 x $10^3$ & 28.7 \\
\hline
Mass window & 6.8 x $10^4$ & 23.8 & 1.1 x $10^3$ & 80.7 \\
\hline
\end{tabular}
\end{center}
\caption{Efficiency of the various cuts applied in preselection (and the mass window cut). Each row shows the
  number of events passing the given cut, as well as all of those preceding it, and the relative efficiency of
  that cut for events which have passed all preceding cuts. All numbers are scaled to the 2012 CMS luminosity
  of 19.5 fb$^{-1}$.}
\label{tab:preselection}
\end{table}

In the final selection step, we look for properties of the jets which can be used to separate signal
from the $b\bar{b}$ background. In general, the jets originating from our signal model have two key
differences from the QCD background: first, they are expected to have more displaced tracks, and for
these tracks to be more highly displaced (as the lifetimes of both the $X$ and the $b$ contribute to
their displacement); and second, we expect the jets to exhibit substructure arising from the presence of
the $b\bar{b}$ pair. After pre-selection, we thus apply stringent requirements on the displaced tracks
and jet substructure for both jets using the variables described in
Sections~\ref{sec:djet} and~\ref{sec:substructure}. The particular values used are as follows:

\begin{enumerate}[\textrm{Final Cut} 1:]
  \item Dijet mass between (80,140) GeV/$c^2$
  \item $\IPS_{4} > 25$ for both jets
  \item $\tau_{21} < 0.65$ for both jets
\end{enumerate}

\begin{figure}[ht]
\centering
\includegraphics[width=0.55\textwidth]{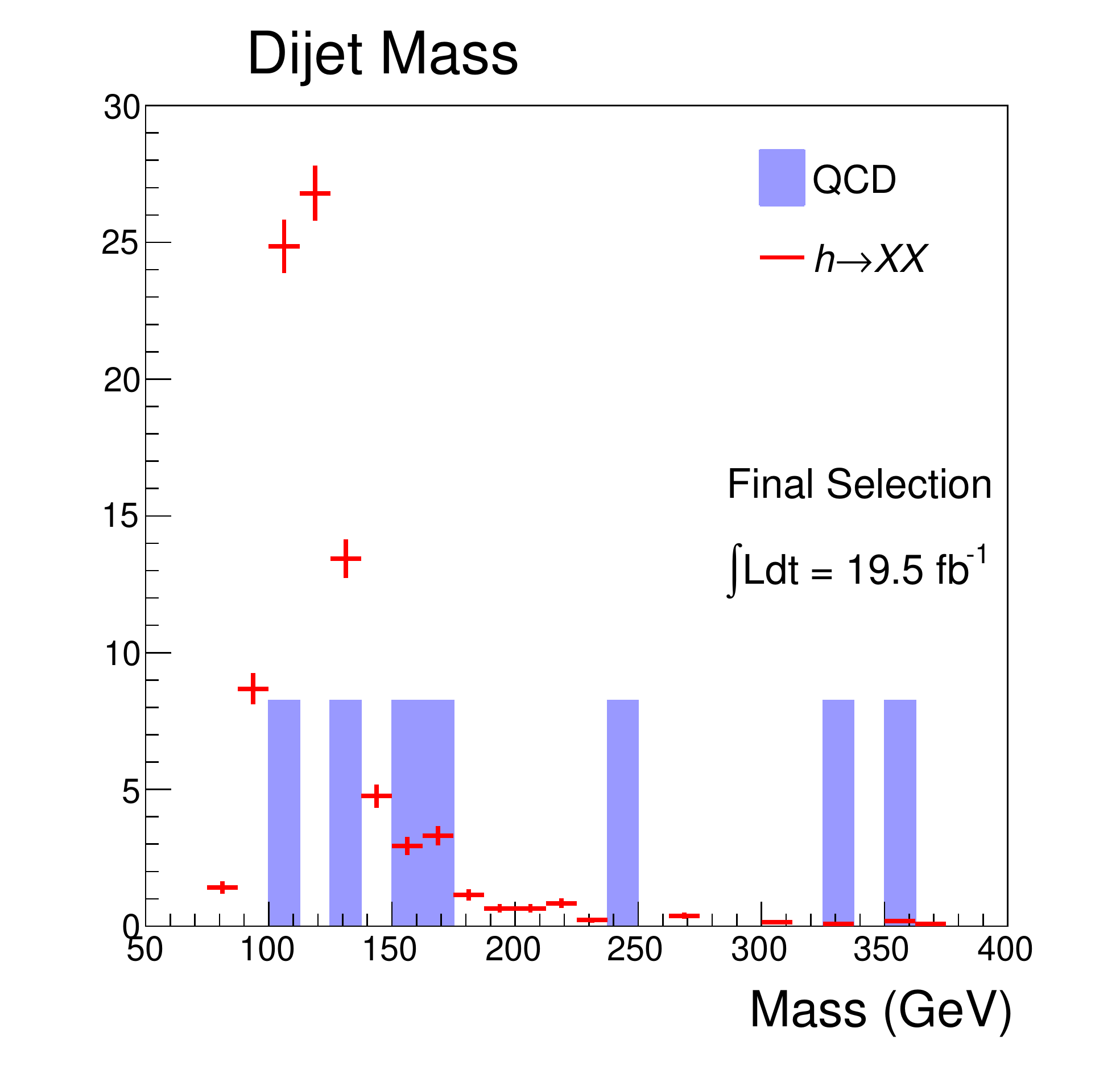}
\caption{Distribution of $m_{jj}$, the invariant
  mass of the two leading jets in the event after final selection. All but the mass window cuts are applied. Shown here are
  simulated signal events and $b\bar{b}$ background, for a signal model with $m_H = 125$ GeV/$c^2$ and $m_X =
  20$ GeV/$c^2$. The signal is clearly visible as a prominent peak over the background.}
\label{fig:finalselection}
\end{figure}

After the final selection, the QCD background is essentially eliminated. Figure~\ref{fig:finalselection} shows the mass distribution in the final signal region (except the mass window cut). However, since our cuts are selecting out sharply falling tails of the QCD background, Monte Carlo simulations can become very unreliable. At the LHC, data driven methods must be employed to obtain a reliable background estimate. We propose such a data-driven estimate in the next section.

\section{Background Estimation}
\label{sec:background}

We adopt a data-driven approach to estimate the expected background, in order to minimize the dependence on
quantities which may not be well-modeled in the Monte Carlo simulation. We use a standard ``ABCD'' approach in
order to estimate the expected amount of background in the signal region. Specifically, we take advantage
of the fact that the two fat jets in an event, as they shower and decay independently, should have uncorrelated
values for the displaced track and substructure variables.

We thus define our ``signal'' region for each individual jet as $\IPS_{4} > 25$ and $\tau_{21} < 0.65$, and define our regions (given events that pass our preselection, including the mass window) as follows:

\begin{itemize}
\item Region A: both jets fail
\item Region B: leading jet passes, second jet fails
\item Region C: leading jet fails, second jet passes
\item Region D: both jets pass (signal region)
\end{itemize}

The final background estimate is then obtained from the ratio $BC/A$. Table~\ref{tab:TnP} shows the results of
applying this technique to the background and signal simulation. We observe that the final estimate is
consistent with the actual number of events in region D in simulation.

\begin{table}[h]
\begin{center}
\begin{tabular}{|r|c|c|c|c|}
\hline
Region & Background & Signal \\
\hline
A (fail/fail) & 6040 $\pm$ 220 & 22 \\
B (pass/fail) & 305 $\pm$ 50 & 47 \\
C (fail/pass) & 345 $\pm$ 53 & 38 \\
D (pass/pass) & 16 $\pm$ 12 & 77 \\
\hline
Final estimate (BC/A) & 17.4 $\pm$ 4.0 & \\
\hline
\end{tabular}
\end{center}
\caption{The number of events in each region for our ABCD technique, scaled to the 2012 CMS luminosity
  of 19.5 fb$^{-1}$. The errors include both the statistical
  uncertainty (from our limited MC sample size) and the systematic
  uncertainty derived from comparison to the data sidebands; the
  latter is the dominant effect.}
\label{tab:TnP}
\end{table}

We can also crosscheck this method in two other ways: first, we can apply the same method but with a
different mass window, in order to obtain a sideband selection of events. Using the background
simulation, we get consistent results using a mass window of (100,160) or (120,180), although the
expected signal in these regions is of course much less.

Another alternative crosscheck is to take advantage of the fact that the $\IPS_{4}$ and $\tau_{21}$ variables
are relatively uncorrelated,\footnote{More specifically, although both of these variables are correlated
through the jet momentum, after the preselection criteria and the dijet mass window cut are applied, the variation in the jet momentum is
reduced, thus decreasing the correlation between these variables arising from the jet $p_T$.} and thus can be
used to define another pair of variables for applying the ABCD method. In this case, the statistics in the
``B'' region are relatively low, thus resulting in a larger systematic uncertainty, so we do not adopt this as
our central estimate. However, we obtain an estimate of $20 \pm 9$ events (statistical uncertainty only),
consistent with our previous estimate.

\section{Results}

\begin{figure}[ht]
\centering
\includegraphics[width=0.6\textwidth]{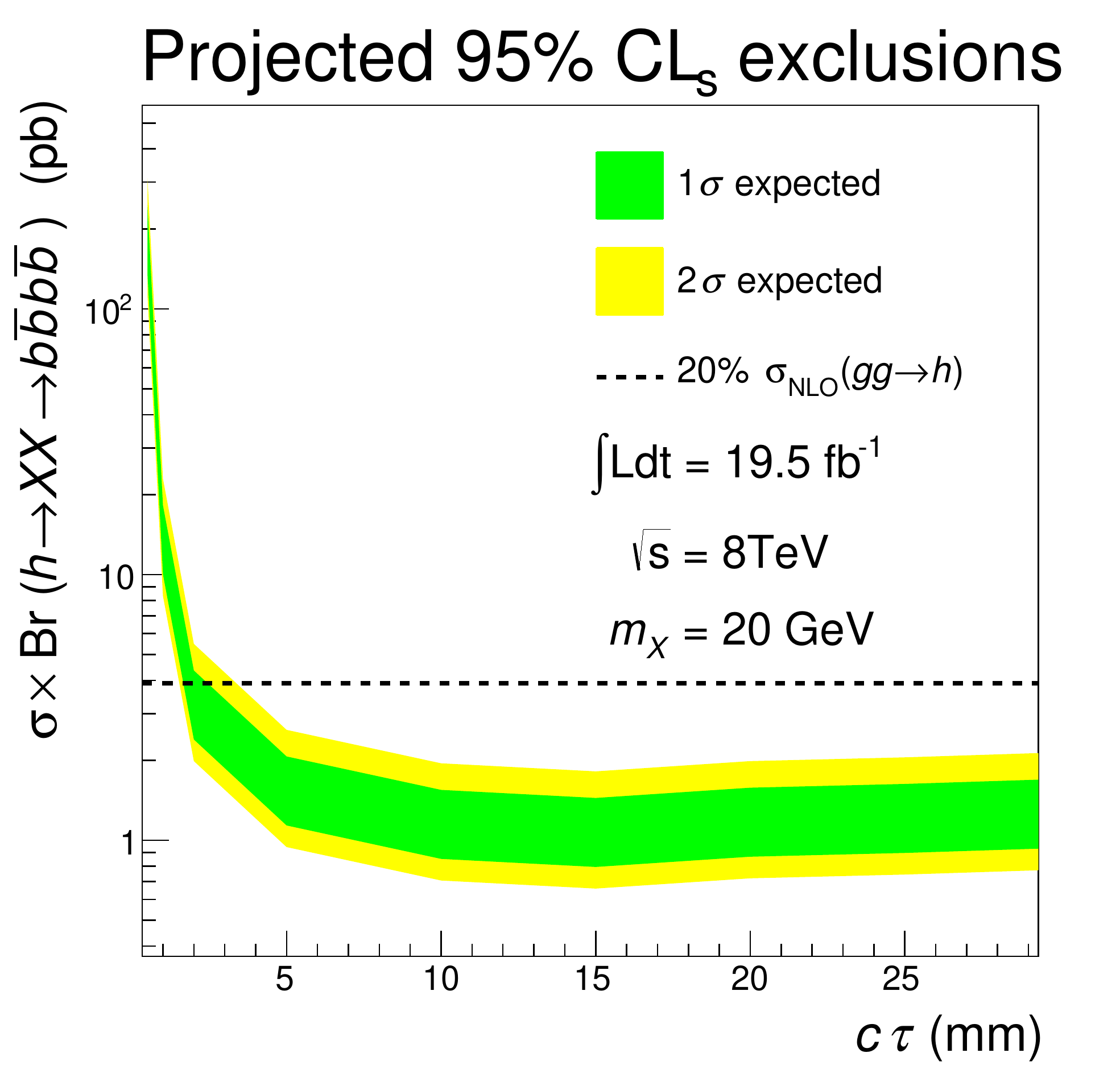}
\caption{Expected limits on the cross-section times branching ratio of the process $h \rightarrow XX
  \rightarrow 4b$ given 19.5 fb$^{-1}$ of data, with 50\% total uncertainty on the background. Systematic uncertainties on the luminosity and efficiency
  are not considered.}
\label{fig:limits}
\end{figure}

Applying the efficiency and the expected QCD background numbers shown in Tables~\ref{tab:preselection}
and~\ref{tab:TnP}, and using the luminosity collected at CMS in 2012 (19.5~fb$^{-1}$), we can set limits on
the cross-section times branching ratio of the Higgs boson to $X\rightarrow b\bar{b}$. The limits are shown in
Figure~\ref{fig:limits} and are obtained using CL$_s$ test statistics and assuming a 50\% total systematic
uncertainty. The systematic uncertainty is conservatively estimated by examining the maximum deviation of the
data-driven method when compared to the actual number of QCD events in different mass windows (which is
limited by our QCD statistics in the higher mass windows). As seen from the figure, for the SM NLO
cross-section and a branching ratio of 20\%, we can exclude down to $c\tau >$ 3~mm for $m_X = $ 20
GeV/$c^2$. The limits for higher $m_X$ are worse due to softer jets and muons. For lower $m_X$, the tracks
become more collimated and the $\tau_{21}$ variable becomes less effective. However, a 20\% $h\rightarrow XX
\rightarrow b\bar{b}b\bar{b}$ branching ratio can be consistently excluded for $m_X\in (15,25)$ GeV/$c^2$ at
$c\tau >$ 3~mm. Further optimization for different mass points may be possible and we leave a detailed study
to the experimental collaborations to properly take detector effects into account.

\begin{figure}[ht]
\centering
\includegraphics[width=0.6\textwidth]{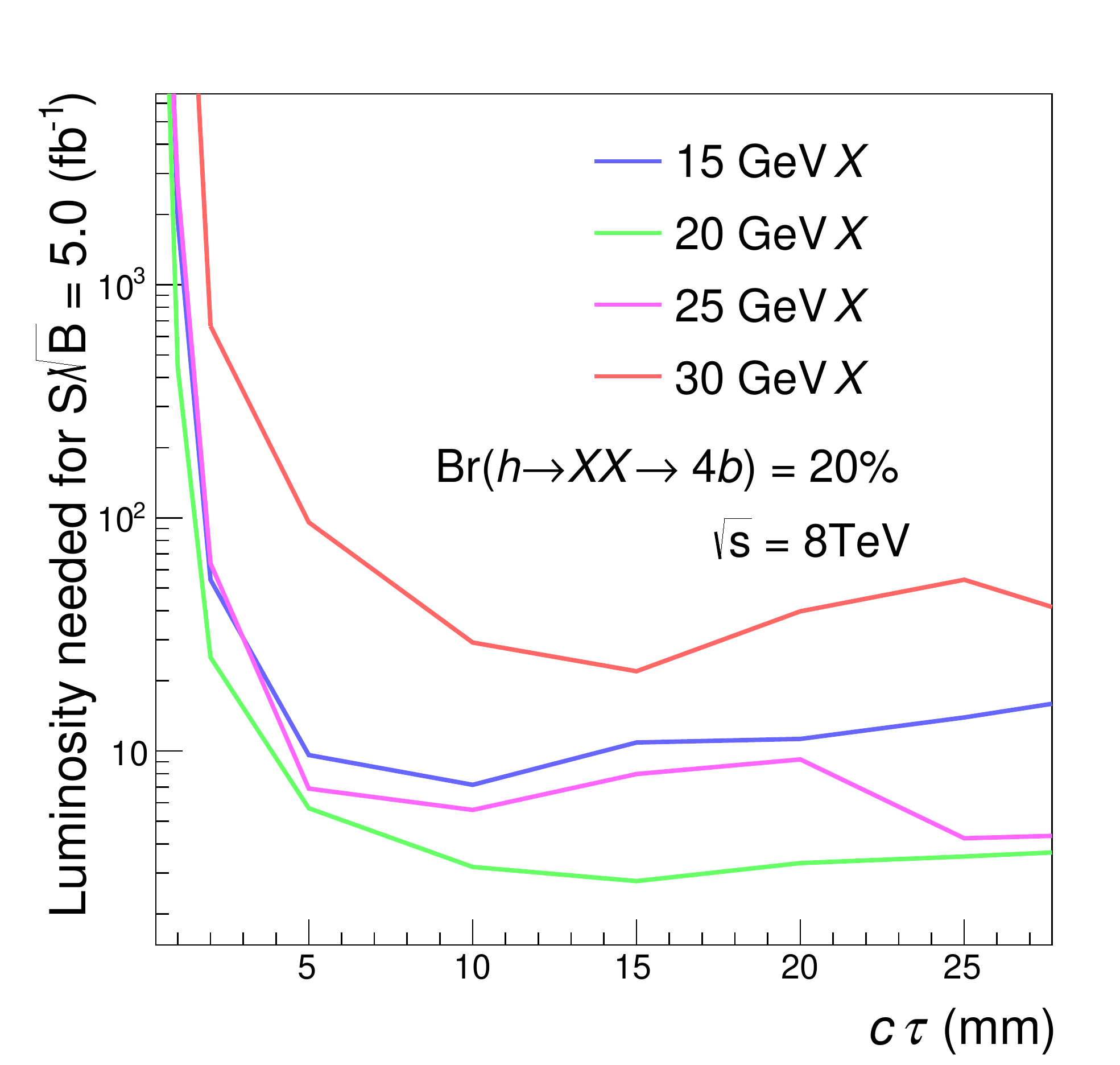}
\caption{Luminosity needed to obtain 5$\sigma$ significance assuming a $20\%$ branching ratio of $h \rightarrow XX
  \rightarrow 4b$. Uncertainities are not included in this plot}
\label{fig:discovery}
\end{figure}

In Figure \ref{fig:discovery} we also consider the discovery potential for $h\rightarrow XX \rightarrow 4b$ decay. For a branching ratio of 20\%, this new decay mode may be discoverable with the current 19.5 fb$^{-1}$ of 8 TeV LHC data for $m_X = $ 15 to 25 GeV/$c^2$. 

\section{Conclusion}

In this note we have introduced several powerful kinematic cuts
designed to discover a Higgs boson decaying to long-lived neutral
particles. The unique features of this channel 
 $h \rightarrow XX \rightarrow 4b$ are two-fold.
First, the highly-displaced
vertices resulting from the decay of the long-lived particles,
some fraction of which occurs before reaching within the beampipe.
The tracks from these vertices will have large IPS.
Second, the long lived particles are boosted enough such the $b\bar{b}$ pairs
are contained within one ``fat jet'', removing combinatoric ambiguities and allowing 
us to take advantage of the jet substructure to distinguish the signal from  the QCD background. 
We have developed a data-driven method to estimate the background and 
for certain values of the (pseudo-)scalar decay length and masses calculated the expected exclusion
(Figure~\ref{fig:limits}) and luminosity needed for discovery (Figure~\ref{fig:discovery}), showing that we
have a strong discovery potential in this channel with about 19.5 fb$^{-1}$ of recorded LHC data. 

Currently, the primarily limiting factor in this analysis is the trigger selection, as the existing triggers
have a relatively low efficiency for the signal considered here. Given the importance of potential new
discovery through exotic topologies that include long-lived particles decaying in the silicon tracker, the
authors were led to evaluate a Graphics Processing Unit (GPU) enhancement of the existing High-Level Trigger
(HLT)~\cite{gpu} to provide new complex triggers that were not previously feasible.  The proposed new
algorithms will allow for the first time the reconstruction of long-lived particles in the tracker system for
the purpose of online selection. New ways of enhancing the trigger performance and the development of
dedicated custom exotic triggers are the key for extending the reach of physics at the LHC.

As a final note, we re-iterate that this study is optimized on a specific mass point, $m_X = 20$ GeV/$c^2$. A
detailed optimization on the cuts on $\IPS_4$ and $\tau_{21}$ could significantly improve the exclusion limits
for different $X$ masses. Furthermore, even though we only focused on the $X \rightarrow b\bar{b}$ case, our
techniques may be sensitive to other channels, such as $X\rightarrow \tau^+ \tau^-$ or a Higgs boson decaying to
a long-lived RPV neutralino $\chi$, which then decays into $\nu b\bar{b}$. With improved triggers for long-lived
particles, additional searches for long-lived $X\rightarrow gg$ and $X\rightarrow qqq$ (for fermionic $X$) may
also be possible. Our search channel also does not have to be limited to a 125 GeV/$c^2$ Higgs particle.  New
particles may potentially be discovered through these long-lived decays.  We leave a detailed optimization of
the different cuts for different decay channels for future work.

\section*{Acknowledgements}

We would like to thank M. J. Strassler  and G. Salam for their critical input to the study presented. 
Their insight and useful discussions helped us clarify some of the theoretical subtleties.
In addition, we would like to thank M. Lisanti for interesting discussions on RPV models and Monte Carlo generators. 
This work is supported by the US Department of Energy, Office of Science Early Career Research 
Program under Award Number DE-SC0003925. H. Lou is supported by the US Department of Energy, Office of Science Graduate Fellowship.

\appendix
\section{Validations}
\label{sec:validation}
\begin{figure}[ht]
\centering
\includegraphics[width=0.6\textwidth]{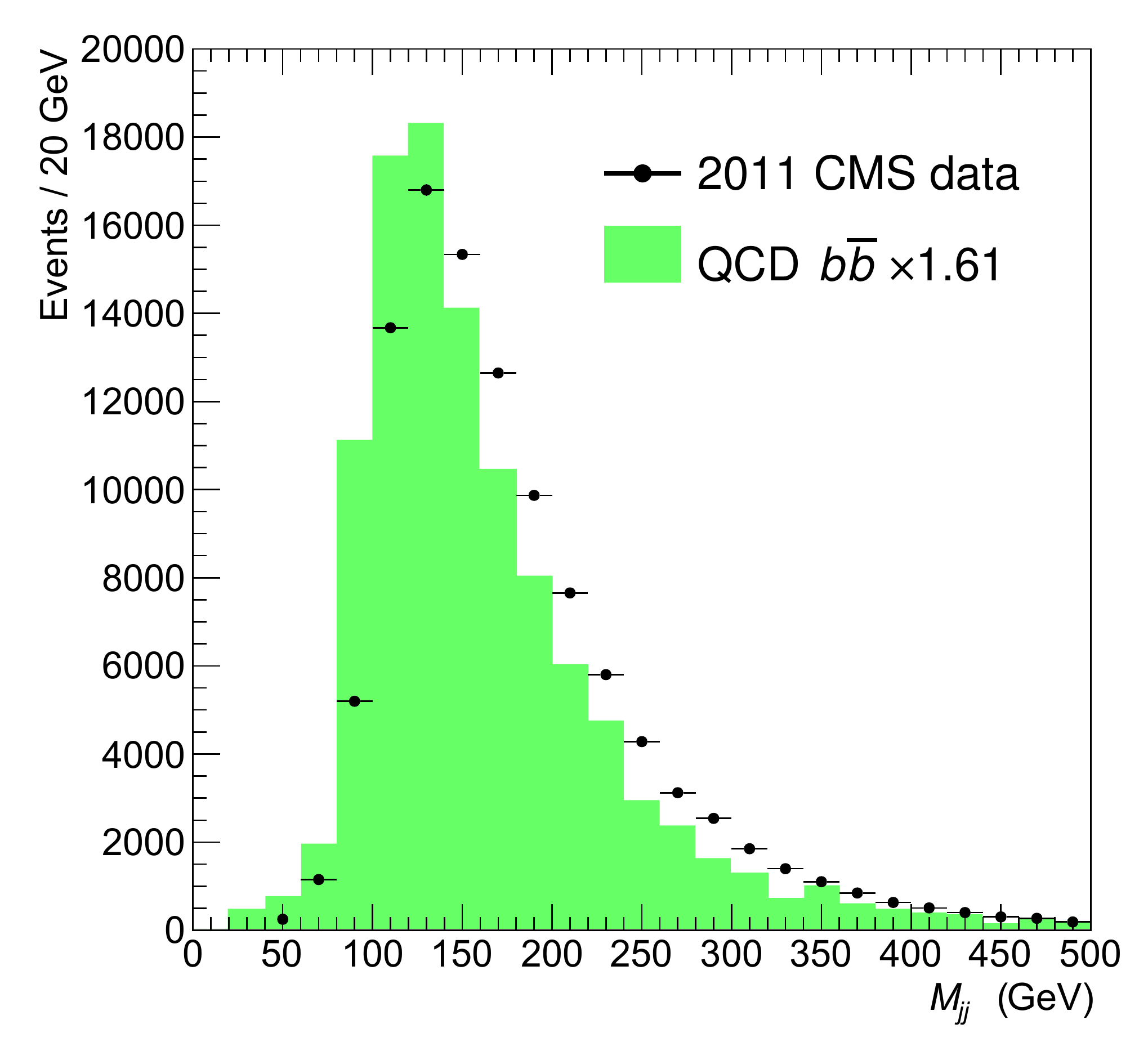}
\caption{The mass distribution of the two leading jets as compared to the CMS analysis in \cite{Chatrchyan:2013qga}.
 A K-factor of 1.6 is obtained and serves as a conservative estimate of $c\bar{c}$ and light flavor contaminations.}
\label{fig:validation_mass}
\end{figure}

In this section, we describe how we obtain a K-factor of 1.6 and validate our $\IPS$ variables against published results. 
A separate 7 TeV $b\bar{b}$ sample is generated for this purpose (matched up to four jets). 500 thousand events were generated using
 \textsc{MadGraph 5} v1.5.7 with the same settings as those listed in Section \ref{sec:generation}. The final state particles are 
clustered using anti-$k_{T}$ algorithm and their momentum smeared with resolution parameters from CMS ~\cite{CMS:2009nxa}. We reproduce
the CMS analysis in \cite{Chatrchyan:2013qga} using the quoted $b$-quark tagging efficiencies and mistag rates. A K-factor of 1.6 is 
obtained by matching our dijet mass distribution against Figure 4a in \cite{Chatrchyan:2013qga}. The resulting distribution is shown 
in Figure \ref{fig:validation_mass}.

To study the effects of fake $b$-tags in events without a true $b$-quark, another 9 million 8 TeV $c\bar{c}$ events were generated
using \textsc{MadGraph 5} v1.5.7 (matched up to 4 jets), and no events passing preselection cuts are found. This suggests that the
rate of fake $b$-tags in our analysis described can be safely neglected. It should be noted that the K-factor obtained from
Figure~\ref{fig:validation_mass} also includes the effects of mistags arising from light-quark contamination; since our
pre-selection cuts are much more stringent than the analysis in \cite{Chatrchyan:2013qga}, using this K-factor should
conservatively include any effects that we might see from fake $b$-tags, so our ignoring of the light flavor QCD background is
justified. The analysis in \cite{Chatrchyan:2013qga} also indicates that vector boson processes contribute less than 1\% of the
total background, hence we neglect their contributions to the background as well.

\begin{figure}[ht!]
\centering
\includegraphics[width=0.6\textwidth]{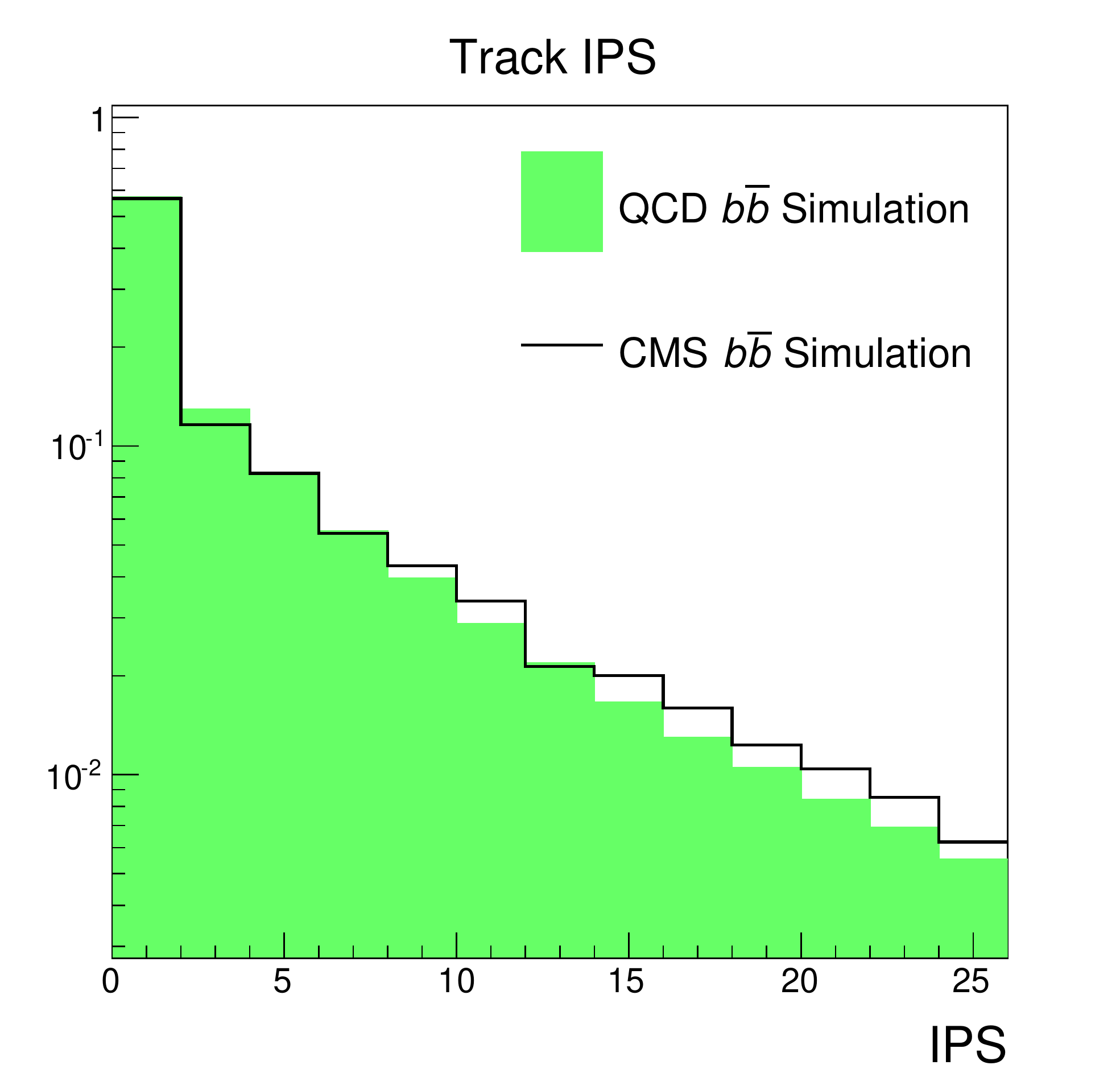}
\caption{The normalized $\IPS$ distribution for $b$-jets when compared to published CMS results in \cite{Rizzi:2006ms}. }
\label{fig:validation_IPS}
\end{figure}

To validate our displaced jet variables outlined in Section \ref{sec:djet}, we also use our 7 TeV validation sample and compare the IPS 
distributions to the published results in Figure 3 in \cite{Rizzi:2006ms}. The normalized distributions are shown in Figure \ref{fig:validation_IPS}.
 Modest disagreements are seen at large $\IPS$. However, the deviations are smaller than our 50\% systematic uncertainty. Given the systematic deviations, 
we also refrain from pushing our $\IPS_4$ cut beyond 25, and our exclusion results are conservative. 

\end{document}